\documentstyle[twocolumn,aps,prl,epsf]{revtex}
\input{epsf}
\input epsf.sty
\begin{document}
\title{The Moving Glass Phase of Dirty Type II Superconductors}
\author{Kyungsun Moon, Richard T. Scalettar, and Gergely T. Zim\'anyi}
\address{Physics Department, University
of California, Davis, CA 95616}
\address{\mbox{ }}
\address{\parbox{14cm}{\rm \mbox{ }\mbox{ }
We study numerically the motion of vortices in dirty type II
superconductors. In two dimensions at strong driving currents,
vortices form highly correlated ``static channels''.
The static structure factor exhibits convincing scaling behaviour,
demonstrating quasi long range translational order
in the transverse direction. However order in the
longitudinal direction is only short range.
We clearly establish the existence of a finite 
transverse critical current, suggesting strong
barriers against small transverse driving forces.
These results are most consistently interpreted in terms of
the recently proposed moving glass picture, modified to account
for the strong anisotropy.
}}\address{\mbox{ }}
\address{\parbox{14cm}{\rm \mbox{ }\mbox{ }
\hskip 5.4cm (30 April 1996)}}
\address{\mbox{ }}
\address{\parbox{14cm}{\rm \mbox{ }\mbox{ }
}}
\maketitle

\narrowtext

Many condensed matter systems reach higher levels of
organization by forming periodic media. Examples range from
crystalline solids to 
Wigner crystals, charge density waves and vortex lattices in 
type II superconductors. A central issue is the effect of
disorder on the stability of such systems. In the case of
flux lattices, for weak disorder it seems 
sufficient to concentrate on elastic deformations
\cite{Giamarchi} which were argued to lead
to a power law decay of lattice correlations, when the periodicity
is properly taken into account. This phase was christened a 
``Bragg-glass" \cite{Giamarchi}. The irrelevancy of topological excitations has been recently confirmed in three dimension, 
while the situation is marginal in two dimension \cite{Hwa}. 
For strong disorder topological excitations such 
as vortex loops can become relevant, giving rise to a Vortex 
glass \cite{Fisher}.
While numerous additional scenarios have been proposed
\cite{Blatter}, there is growing experimental evidence
supporting the basic picture of two types of glasses as the 
disorder or the magnetic field is increased \cite{Gammel,Koch}.

Upon increasing the external force beyond a certain depinning strength,
these periodic (or glassy) media become mobile. Early work developed
perturbation studies at high velocities $v$ in powers of
$1/v$\cite{Larkin,Schmid}.
Recently Koshelev and Vinokur argued that the effect of the random
potential is seriously weakened at high velocities, as it 
``averages out", thus reestablishing the long range solid order
\cite{Koshelev}. Their numerical simulations suggested that this change
occurs abruptly, giving rise to a genuine dynamic phase transition. 
On the other hand Giamarchi and Le Doussal pointed out that 
some components of the disorder remain unsuppressed, as they represent 
{\em static} perturbations\cite{LeDoussal}. 
These destroy the moving solid, and 
stabilize a glassy phase instead with quasi-long range order (QLRO)
only, giving rise to a moving glass. 
Balents and Fisher, in their study of the 
related charge density wave systems also find that the moving phase 
possesses QLRO only, but may nevertheless support true long 
range temporal order\cite{Balents}.
The physics of the moving glass phase is\cite{LeDoussal} that the 
vortices form highly correlated {\it static channels}. This picture 
leads to a logarithmic decay at large distances for the displacement
correlations. It also gives rise to diverging potential barriers 
and consequently a finite critical current against an additional
transverse current.

In this Letter, we report a detailed numerical study
of the moving glass phase. We establish that the basic picture of
``static channels" is indeed characteristic of the system.
By studying the static structure factor we demonstrate the existence
of quasi long range translational order with an algebraic decay 
in the transverse direction but find only  
short range longitudinal order, giving rise to a very anisotropic
glass. As a direct consequence we predict the absence of a narrow
band noise for moving glasses in the 2D vortex systems.
Finally the existence of a critical transverse current will be clearly
established.

We employ overdamped Molecular Dynamics 
(MD) simulations at zero temperature to study two dimensional 
interacting vortices in the presence of point disorder,
\begin{equation}
\gamma \frac {d{\bf r}_i} {dt} = \sum_{j\ne i} {\bf F}_v({\bf r}_i-{\bf r}_j)  
+ \sum_{j} 
{\bf F}_{pin}({\bf r}_i-{\bf R}_{j}) + {\bf F}_{L}~~.
\end{equation}
Here $\gamma$ is the damping parameter, ${\bf R}_j$ specifies 
the pinning center positions, ${\bf r}_{i}$ denotes the location
of the {\em i-th} vortex, and ${\bf F}_L$ is the Lorentz force, 
exerted by the external driving current.   
The force between vortices is given by 
\begin{equation}
{\bf F}_v ({\bf r})=F_0 (1 - {\tilde r}^2)^2 \;\frac {{\tilde {\bf r}}}
{{\tilde r}^2},
\end{equation}
where ${\tilde r}=r/R_{cut},  
F_0=V_{0}/R_{cut}$, 
and we choose $R_{cut}=3.6 a_0$, where $a_{0}$ is the mean vortex
spacing. Here $\gamma, a_0$, and $V_{0} (\cong \Phi_0^2/8\pi^2\lambda^2)$ 
define the units of time, length and energy, respectively.
The pinning force is taken as
\begin{equation}
{\bf F}_{pin}({\bar r})=-4 F_p (1-{\bar r}^2) \; {\bar {\bf r}} ~~. 
\end{equation}
Here ${\bar r}=r/R_{pin}$ and $R_{pin}=0.25 a_0$.
We worked with a fixed density of pinning centers of $5.77/a_0^2$.
This set of parameters was chosen to optimize the
convergence of the numerical procedure.

Now we construct the phase diagram in the driving force -
pinning strength plane, at zero temperature.
With increasing driving currents three phases emerge: 
a pinned glass, a plastic flow regime, and some kind of an ordered phase. 
At low forces the vortices remain pinned, forming a glassy phase. 
As the Lorentz force is increased beyond a critical value
$F_{d}$, the vortices depin. This transition, and in particular the value of $F_d$ can be well captured by studying the current-voltage (IV)
characteristics. The resulting values of $F_d$ were used to construct the lower phase boundary in Fig.1.
For strong disorder, just above $F_{d}$ vortices form a pattern of 
pinned and unpinned regions, often described as ``plastic flow''
\cite{Blatter}. In this regime $F_{d}$ scales linearly with the pinning strength, whereas for weak disorder the relation is quadratic. 
The near-linearity of the phase boundary in Fig.1 indicates
that we concentrated on the regime of strong disorder.

\begin{figure}[t]
\epsfxsize=3.in
\epsfysize=3.in
\epsffile{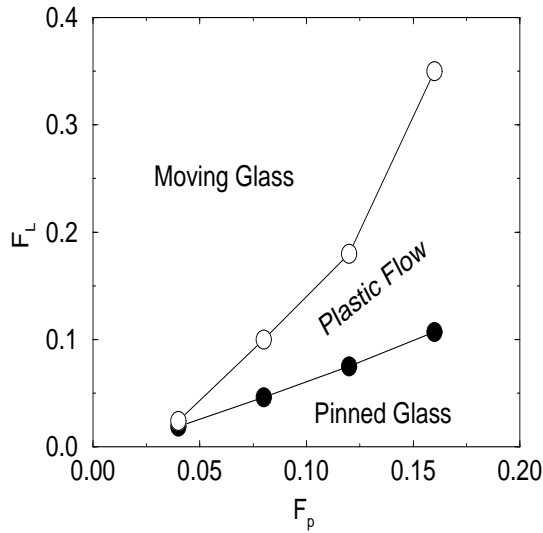}
\vskip 0.5cm
\caption{The dynamic phase diagram. $F_p$ is the pinning strength,
$F_L$ is the Lorentz force. Open circles represent $F_g$, solid circles
$F_{d}$.}
\label{figure1}
\end{figure}

Upon further increase of the driving force, Koshelev and Vinokur 
argued\cite{Koshelev} that the effects of the disorder on the lattice
displacements ``average out'', leading to solid ordering in 3D,
and quasi long range translational order in 2D.
Alternatively, Giamarchi and Le Doussal suggested\cite{LeDoussal} 
that the vortex system forms a moving glass.
To distinguish between these propositions we first determine
the location of the phase boundary, then explore the physics
of the high-velocity phase.

The phase boundary $F_g$ between the plastic flow regime and the 
putative moving solid can be established by measuring
the static structure factor $S({\bf k})$.
In the plastic flow regime the absence of ordering
manifests itself in a central peak and a structureless ring (see
lower panel in Fig.2).
In the high velocity phase $F_{L}>F_g$, one expects to see six-fold
coordinated Bragg peaks, if a moving solid is formed.  We do
indeed observe a sharp transition into a phase with 
well developed peaks, however the peak pattern is strongly
anisotropic (upper panel in Fig.2).

\begin{figure}[t]
\epsfxsize=3.in
\epsfysize=5.in
\epsffile{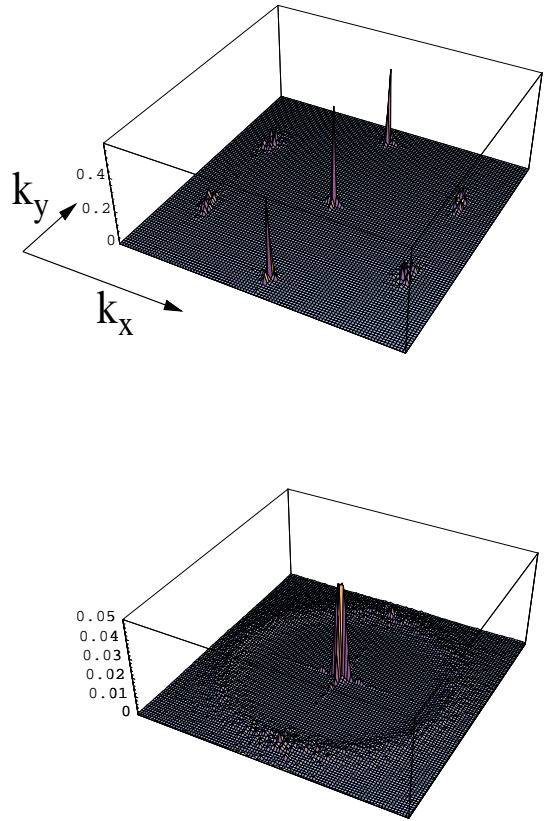}
\vskip 0.5cm
\caption{The static structure factor $S({\bf k})$.
$F_L=0.6$ in the upper and $0.2$ in the lower panel.
The number of vortices is $960$ and the disorder strength
$F_p=0.16$. The critical force $F_g$ indicating the transition from the
plastic flow regime to the moving glass is about $0.35$.}
\label{figure2}
\end{figure}

The upper phase boundary in Fig.1 was determined by mapping out $F_g$ 
for several disorder strengths. One can see that for strong disorder
indeed all three expected phases are observed, whereas for weak disorder
there is no robust evidence for an intervening plastic
flow regime. Either that phase occupies a very slim region
in parameter space, or there is a direct pinned Bragg glass-to-moving
glass transition. This transition is much
harder to identify because both phases exhibit quasi long range order,
and thus the structure factors are very similar in the two phases.

The central issue of our paper is to elucidate whether
at high velocities the system supports a moving solid or a moving glass.
To address this issue we first analyse $S({\bf k})$,
\begin{equation}
S({\bf k})=  \; {1\over L^d}
\sum_{i,j} e^{i {\bf k}\cdot ({\bf r}_i (t)-{\bf r}_j (t))}.
\end{equation}
The pinning strength $F_p$ is fixed to be 0.16 and the applied force 
$F_L = 0.6$  is well above the corresponding 
critical force $F_g \cong 0.35$.
We simulate five different system sizes 
with fixed vortex density and 
number of vortices ranging from 240 to 1500.
The initial configurations are chosen randomly. We let the MD simulations
evolve with time, make sure that the system reaches its steady state,
then freeze the vortex configuration and measure $S({\bf k})$.
In the steady state the vortices form an orderly
array. Its principal lattice vector in most cases is 
aligned with the direction of motion.
It was argued that the system chooses such an orientation
to minimize the power-dissipation\cite{Schmid}.
However the details of this alignment have yet to be understood.
The peaks at the reciprocal lattice vectors {\it perpendicular} 
to the motion exhibit quite convincing power law behavior. The 
height of the other peaks however decay very rapidly with increasing
system sizes, suggesting that the corresponding correlations are 
short ranged. 

\begin{figure}[t]
\epsfxsize=3.in
\epsfysize=3.in
\epsffile{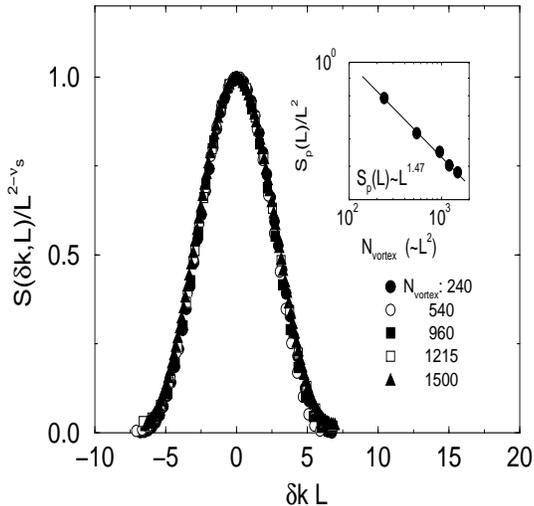}
\vskip 0.5cm
\caption{Finite size scaling of the static structure factor $S({\bf k})$.  
The driving force $F_L = 0.6$ is well above the critical
force $F_g \sim 0.35$.
The inset shows power law dependence of the peak heights
with varying system size; $S_p(L)\sim L^{2-\nu_s} $ with
$\nu_s\sim 0.53\pm 0.1$.}
\label{figure3}
\end{figure}

To study the peaks of $S({\bf k})$ at the reciprocal 
lattice vector ${\bf G}_0=(0, \pm 4\pi/\sqrt{3})$,
we write down the following finite size scaling form:  
\begin{equation}
S(\delta k,L) = L^{d-\nu_s} G(\delta k\, L),
\end{equation}
with $\delta k=|{\bf k}-{\bf G}_0|$.  
In Fig.(\ref{figure3}) the scaling function $G(x)$ 
is plotted with respect to the dimensionless scaling variable 
$x=\delta k L$ for the five system sizes. The peak amplitude scales 
with the system size as $L^{2-\nu_s}$ with $\nu_s=0.53\pm 0.1$, 
as shown in the inset of Fig.(\ref{figure3}). 
Using this value of $\nu_s$ the normalized structure factor exhibits a 
convincing data collapse onto a single curve. 
This confirms the scaling behaviour 
$S({\bf k}) \cong |{\bf k}-{\bf G}_0|^{-1.47}$
around the peaks.

This implies that phase slips in the transverse direction are 
strongly suppressed, justifying the elastic approach.
In contrast, the peaks at momenta with nonzero longitudinal components 
decay rapidly with system size. Several physical mechanisms can 
lead to such a decay. Extensive study of single snapshots of the spatial
distribution of vortices
suggests that {\it phase slips} between longitudinal boundaries
of elastic domains are primarily responsible.
This observation questions the elastic theory for this direction. 
Clearly a more complete understanding is needed
on this issue, especially in two dimension\cite{comment1}.

A measurable consequence of the absence of translational order
in the longitudinal direction should be the corresponding
absence of narrow band noise in 2D driven vortex systems.
The same conclusion was reached for the analogous CDW models in 2D 
in Ref.\cite{Balents}.

\begin{figure}[t]
\epsfxsize=3.in
\epsfysize=3.in
\epsffile{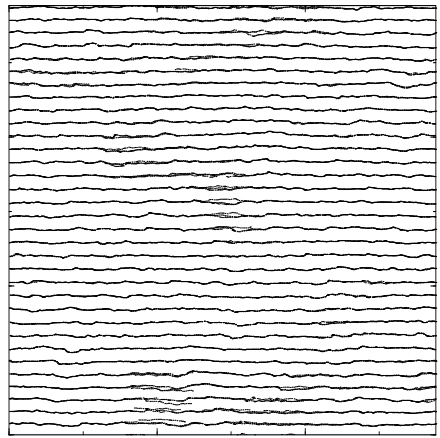}
\vskip 0.5cm
\caption{The static channels in the steady state.
The small fluctuations transverse to the channels are due to the
perturbation from the rapidly varying time-dependent component of
the disorder, viewed in the rest frame of vortices.}
\label{figure4}
\end{figure}

The scaling behaviour of $S({\bf k})$ does not distinguish between
a quasi-solid and a moving glass. The proposition of the glassy phase
rests on the argument that certain components of the disorder
{\it do not} average out, but present a {\it static} perturbation.
If so, the moving vortices should form static channels, which do not
change their shape with time. To study this we now map out the
trajectories of the vortices. Making sure that the flow reached its 
steady state, we take a large number of consecutive snap shots, which 
are then displayed on top of each other. The resulting 
Fig.(\ref{figure4}) clearly demonstrates the formation of static channels.
The very existence of these time-independent channels lends 
strong credence to the moving glass picture.

The moving glass picture also implies the existence of diverging
barriers against small transverse currents, leading to
a {\it finite transverse critical current}.
This critical current is the largest when one of the principal lattice
vectors is parallel to the motion \cite{LeDoussal}.

\begin{figure}[t]
\epsfxsize=3.in
\epsfysize=3.in
\epsffile{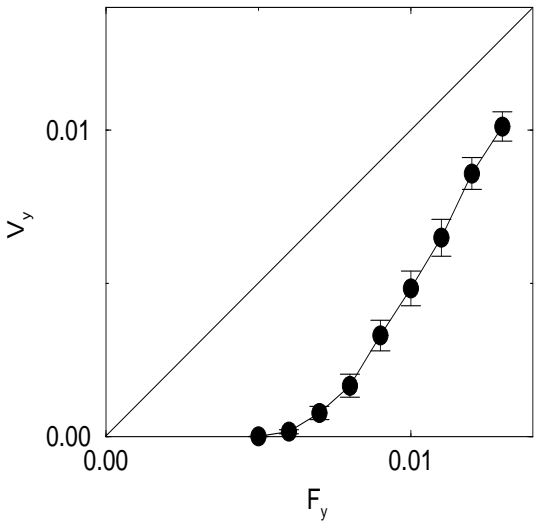}
\vskip 0.5cm
\caption{IV characteristics, showing transverse
velocity $v_y$ as a function of transverse force $F_y$.
Here $F_L=0.6$. The existence of a finite critical
force $F^y_d \sim 0.006\pm 0.001$ is clear. 
The straight line represents a free flux flow response.}
\label{figure5}
\end{figure}

We select 50 disorder realizations which lead to a steady state
with one of its primitive lattice vectors parallel to the direction of
the velocity. After the system reaches the steady state,
a small additional transverse force is applied 
and the transverse velocity  $v_y$ is measured. 
Fig.(\ref{figure5}) clearly exhibits
a finite critical transverse force $F^y_d \cong 0.006\pm 0.001$.
$F^y_d$ is much smaller than the longitudinal critical  
force $F_d \cong 0.1$, since the effective disorder strength is 
inversely proportional to the velocity of the moving glass
\cite {LeDoussal}. Simulations for several different system sizes 
result in identical critical currents, indicating that
this threshold behaviour is not a finite size effect\cite{comment2,Kmoon}.

In sum, we explored the moving phase of vortex systems at high velocities.
We measured the IV characteristics to determine the critical 
depinning force $F_{d}$, establishing the phase boundary
between the pinned glass and the plastic flow regime.
Next we studied the structure factor $S({\bf k})$, which exhibited
a sharp transition from its ring shape in the plastic flow regime
into a phase with an anisotropic peak pattern.
The Bragg peaks suggest the existence of power law order in the 
transverse directions, but only short range longitudinal order.
Thus we expect the absence of a narrow band noise. 
We demonstrated the formation of static channels,
and found a finite critical transverse current.
These results are consistent with the suggestion that the 
driven vortex system in two dimension forms an 
{\em anisotropic moving glass}.

We acknowledge useful discussions with T. Giamarchi, V. Vinokur, 
L. Balents, A, Koshelev and P. LeDoussal. 
K. Moon wishes to thank S. M. Girvin for generously allowing 
access to his computing facilities.
This work has been supported by NSF-DMR-95-28535 and by the LACOR
program of the Los Alamos National Laboratory.

\end{document}